\documentclass[twocolumn, twocolappendix]{aastex631}

\usepackage{acronym}
\usepackage{xspace}
\usepackage{amsmath}
\usepackage{hhline}

\newacro{bbh}[BBH]{binary black hole}
\newacro{BH}[BH]{black hole}
\acrodef{EoS}[EoS]{Equation of State}
\acrodef{BHNS}[BHNS]{black hole--neutron star binary}
\acrodef{NSBH}[NSBH]{neutron star--black hole}
\acrodef{NS}[NS]{neutron star}
\newacro{bns}[BNS]{binary neutron star}
\acrodef{FAR}[FAR]{false alarm rate}
\newacro{bf}[BF]{Bayes' factor}
\newacro{cbc}[CBC]{compact binary coalescence}
\acrodef{CE}[CE]{Common Envelope}
\acrodef{SNe}[SNe]{Supernova}
\newacro{da}[DA]{data analysis}
\newacro{et}[ET]{Einstein Telescope}
\newacro{eob}[EOB]{Effective-One-Body}
\newacro{fd}[FD]{frequency domain}
\newacro{GW}[GW]{gravitational-wave}
\newacro{gr}[GR]{general relativity}
\newacro{hm}[HM]{Higher mode}
\newacro{ifo}[IFO]{interferometer}
\newacro{imr}[IMR]{inspiral-merger-ringdown}
\newacro{im}[IM]{inspiral-to-merger}
\newacro{kagra}[KAGRA]{Kamioka Gravitational Wave Detector}
\newacro{ligo}[LIGO]{Laser Interferometer Gravitational-Wave Observatory}
\newacro{lso}[LSO]{Last Stable Orbit}
\newacro{lvc}[LVC]{LIGO-Virgo Collaboration}
\newacro{lvk}[LVK]{LIGO-Virgo-Kagra Collaboration}
\newacro{lo}[LO]{leading order}
\newacro{ns}[NS]{neutron star}
\newacro{nr}[NR]{numerical relativity}
\newacro{pn}[PN]{post-Newtonian}
\newacro{pe}[PE]{parameter estimation}
\newacro{psd}[PSD]{power spectral density}
\acrodef{PSD}[PSD]{power spectral density}
\acrodef{KN}[KN]{kilonova}

\acrodefplural{KN}[KNe]{kilonovae}
\newacro{qc}[QC]{quasi-circular}
\newacro{snr}[SNR]{signal-to-noise ratio}
\acrodef{SNR}[SNR]{signal-to-noise ratio}
\newacro{ng}[NG]{Next Generation}
\newacro{eos}[EoS]{Equation of State}

\definecolor{cyan}{rgb}{0,0.9,0.9}
\definecolor{orange}{rgb}{0.9,0.5,0}
\definecolor{magenta}{rgb}{1,0,1}
\definecolor{purple}{rgb}{0.8,0.4,0.8}
\definecolor{gray}{rgb}{0.8242,0.8242,0.8242}
\definecolor{dodgerblue}{rgb}{0.12, 0.56, 1.0}

\newcommand{\BILBY}{\textsc{Bilby}\xspace}

\newcommand{\BAYESLINE}{\textsc{BayesLine}\xspace}
\newcommand{\DYNESTY}{\textsc{Dynesty}\xspace}

\begin{document}

\title{Everything everywhere all at once: A detailed study of GW230529}

\correspondingauthor{kbc595@psu.edu}

\author[0000-0003-4750-5551]{Koustav Chandra}
\affiliation{Institute for Gravitation \& the Cosmos and Physics Department, The Pennsylvania State University, University Park PA 16802, USA}

\author[0000-0001-6932-8715]{Ish Gupta}
\affiliation{Institute for Gravitation \& the Cosmos and Physics Department, The Pennsylvania State University, University Park PA 16802, USA}

\author[0000-0001-7239-0659]{Rossella Gamba}
\affiliation{Institute for Gravitation \& the Cosmos and Physics Department, The Pennsylvania State University, University Park PA 16802, USA}
\affiliation{Department of Physics, University of California, Berkeley, CA 94720, USA}

\author[0000-0002-5700-282X]{Rahul Kashyap}
\affiliation{Institute for Gravitation \& the Cosmos and Physics Department, The Pennsylvania State University, University Park PA 16802, USA}

\author[0000-0001-5867-5033]{Debatri Chattopadhyay}
\affiliation{School of Physics and Astronomy, Cardiff University, Cardiff, CF24 3AA, United Kingdom}

\author[0000-0002-5034-9353]{Alejandra Gonzalez}
\affiliation{Theoretisch-Physikalisches Institut, Friedrich-Schiller-Universit{\"a}t Jena, 07743, Jena, Germany}

\author[0000-0002-2334-0935]{Sebastiano Bernuzzi}
\affiliation{Theoretisch-Physikalisches Institut, Friedrich-Schiller-Universit{\"a}t Jena, 07743, Jena, Germany}

\author[0000-0003-3845-7586]{B.S. Sathyaprakash}
\affiliation{Institute for Gravitation \& the Cosmos and Physics Department, The Pennsylvania State University, University Park PA 16802, USA}
\affiliation{Department of Astronomy and Astrophysics, Penn State University, University Park PA 16802, USA}
\affiliation{School of Physics and Astronomy, Cardiff University, Cardiff, CF24 3AA, United Kingdom}

\begin{abstract}

This study investigates the origins of GW230529, delving into its formation from massive stars within isolated binary systems. Utilizing population synthesis models, we present compelling evidence that the \acl{NS} component forms second. However, the event's low \acl{SNR} introduces complexities in identifying the underlying physical mechanisms driving its formation. Augmenting our analysis with insights from numerical relativity, we estimate the final black hole mass and spin to be approximately $5.3$ $M_\odot$ and $0.53$, respectively. Furthermore, we employ the obtained posterior samples to calculate the ejecta mass and kilonova light curves resulting from r-process nucleosynthesis. We find the ejecta mass to range within $0$--$0.06$ $M_{\odot}$, contingent on the neutron star equation of state. The peak brightness of the kilonovae light curves indicates that targeted follow-up observations with a Rubin-like observatory may have detected this emission.

\end{abstract}

\section{introduction}

Electromagnetic observations have played a pivotal role in constraining the mass spectrum of \acp{NS} and \acp{BH}. Notably, X-ray and radio observations have consistently indicated that the maximum mass of \acp{NS} falls within the range of $2-2.6 M_{\odot}$, while \ac{BH} masses have been found to exceed $5 M_{\odot}$~\citep{Antoniadis:2016hxz, Alsing:2017bbc, Farr_2020, Fonseca:2021wxt, Romani:2022jhd, Bailyn:1997xt, Ozel:2010su, Farr:2010tu}. Consequently, these detections have implied a notable absence of compact binaries within the mass range of $2.6-5 M_{\odot}$, a notion reinforced by initial \ac{GW} observations \cite{LIGOScientific:2018mvr, LIGOScientific:2020ibl}. However, \acf{SNe} simulations have predicted the existence of \acp{BH} within this gap~\citep{Fryer:2012}.
Furthermore, the recent observation of the binary merger GW230529\_181500 (hereafter referred to as GW230529) has provided conclusive evidence for the existence of compact objects within this mass range~\citep{LVK:2024elc}.

This event was observed only in LIGO Livingston ~\citep{LIGOScientific:2014pky}, with a \acf{SNR} $\gtrsim 11$ and a low \ac{FAR} of less than one event per thousand years. Assuming GW230529 is the result of the merger of a \acl{BHNS} on a quasi-circular orbit, the data are consistent with the merger of two compact objects with masses $3.6^{+0.8}_{-1.2}M_\odot$ and $1.4^{+0.6}_{-0.2}M_\odot$ (90\% credible intervals), making it the most symmetric mixed compact binary merger detected via \acp{GW}.

\begin{figure*}[htb]
    \centering
 
     \includegraphics[width=\textwidth]{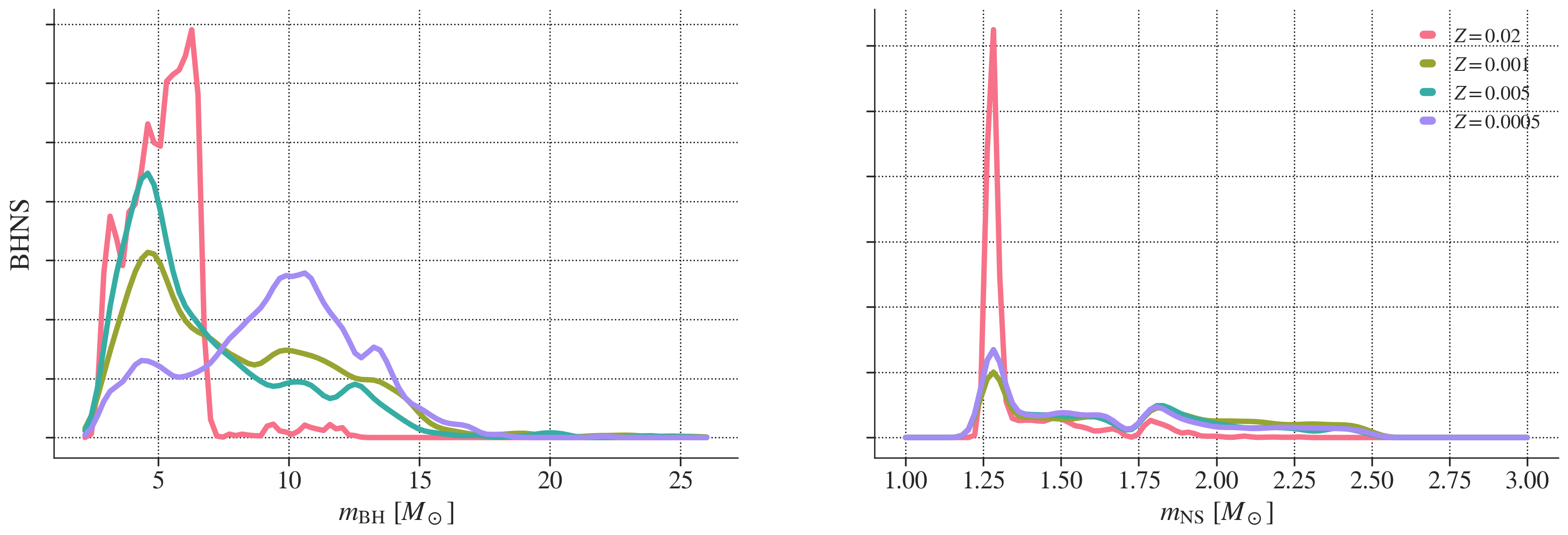}
    \includegraphics[width=\textwidth]{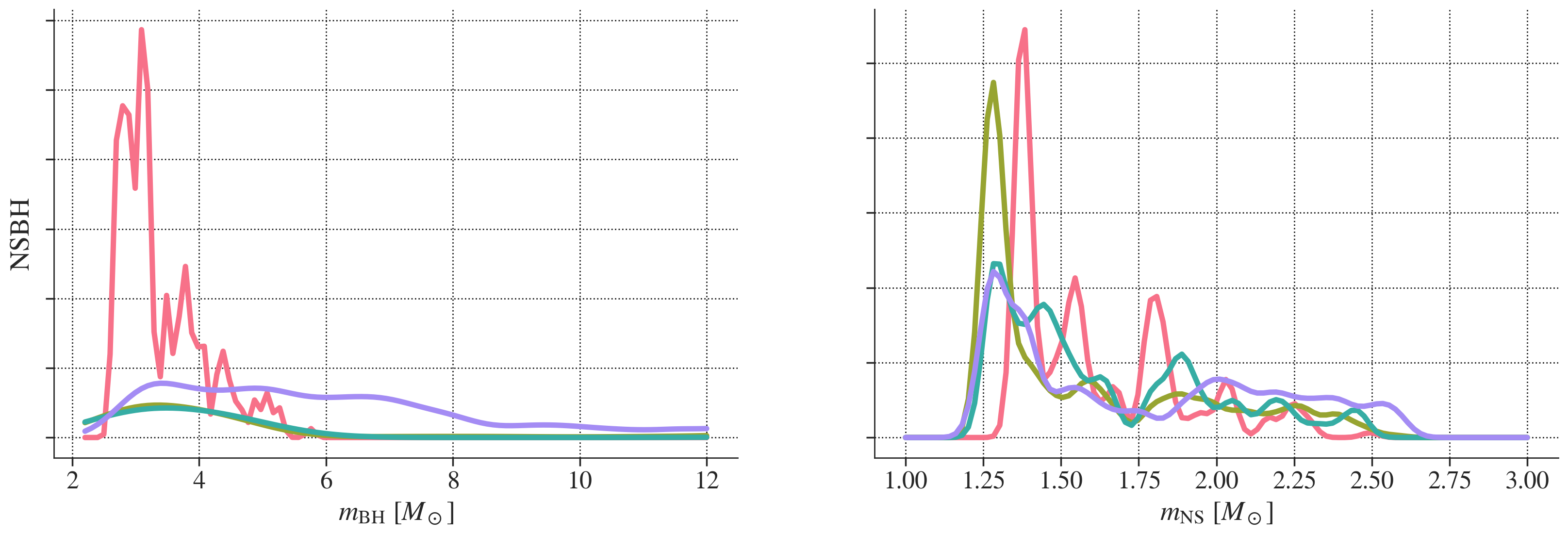}
    \caption{Predicted component mass distributions for \ac{BHNS} (top panel) and \ac{NSBH} systems (bottom panel) across varying sub-solar metallicities, as outlined in \citet{Chattopadhyay:2020lff}. As can be observed, \ac{NSBH} systems tend to produce binaries with lighter \acp{BH} across different metallicity.}
    \label{fig:bhns-prior-metallicity}
\end{figure*}

Measuring these mass parameters is part of the inverse problem. Other parameters of astrophysical interest include, but are not limited to, the components' spins, the tidal properties and the remnant properties. A Bayesian inference approach is typically employed, which necessitates model evaluations 
to reliably infer the posterior distribution for the parameters that characterize the observed signal. While the data containing GW230529 was matched against several state-of-art \ac{GW} signal models, \citet{LVK:2024elc}'s initial analysis assumed priors that are effectively flat in component masses and spin magnitudes. These astrophysically-agnostic priors do not provide information about GW230529's possible formation mechanism.

Therefore, \citet{LVK:2024elc} reweighed the posterior samples using a few astrophysically-informed mass and spin distributions to constrain the source properties~\citep{Fishbach:2020ryj, Farah:2021qom, Ray:2023upk}. They unequivocally observed that the component masses inferred for GW230529 differ across different population-informed priors, demonstrating that the choice of prior significantly impacts the inferred masses and spins of the GW230529's source. This is expected, given the event's low SNR. Therefore, it is necessary to determine whether the data support a given model, as strong priors might drive the posteriors to arbitrary values at the expense of Bayesian evidence.

Also, the employed astrophysically-informed priors are phenomenological and contain only an \textit{observationally constrained} understanding of stellar and binary evolution. Further,~\citet{Zhu:2024cvt} found that the GW230529's masses are close to those predicted in their \ac{BHNS} population simulations.

In this article, we re-analyze the data containing GW230529 with population-informed priors to provide conclusive insights into the probable physical processes underlying this binary formation. Leading formation models of GW230529-like systems include the isolated evolution of massive binary stars in galactic fields via the \acf{CE} and the dynamical assembly aided by either a tertiary companion, multiple exchanges in dense clusters or gas-assisted migration (see~\cite{Mandel:2021smh} and references therein). While direct collapse leading to the formation of the primary component is unlikely due to its low mass, recent population synthesis models have argued the plausibility of such systems arising from isolated evolution scenarios~\citep{Chattopadhyay:2020lff, Broekgaarden:2021iew, Broekgaarden:2021b, Chattopadhyay:2022}. However, these models have major uncertainties such as mass loss, mass transfer and the impact of supernova explosions, resulting in a broad spectrum of merger rate predictions and varying mass and spin distributions. 

We describe our methodology in Sec.~\ref{sec:methods}, utilizing it subsequently in Sec.~\ref{sec:bhns-nsbh} to ascertain the nature of the merger—specifically, whether the \ac{NS} or \ac{BH} formed first. Consequently, in Sec.~\ref{sec:priors}, we employ the evidence ratio of posterior samples and the distribution of log-likelihood ratios to distinguish between various formation models, shedding light on the most likely formation mechanism of GW230529.

We then use the posterior samples corresponding to these models to analyse the characteristics of the remnant \ac{BH}, as detailed in Sec.~\ref{sec:remnant-mma}. Our analysis is conducted using a selection of \acp{EoS} that adhere to current astrophysical and nuclear constraints~\citep{Breschi:2024qlc}. Furthermore, this section assesses the probability of \ac{NS}'s tidal disruption during the merger.

Subsequently, in Sec.~\ref{sec:ejecta}, we present the range of ejecta masses that inform the r-process synthesis resulting from such events. Our analysis incorporates four fiducial \acp{EoS}, namely \texttt{APR4}~\citep{Akmal:1998cf}, \texttt{SLy} ~\citep{Chabanat:1997un, Douchin:2001sv}, \texttt{DD2}~\citep{Typel:2009sy, Hempel:2009mc} and \texttt{H4} \citep{Glendenning:1991es, Lackey:2005tk, Read:2008iy}.

Finally, in Sec.~\ref{sec:lightcurves}, we compute the lightcurves associated with \acp{KN} for the event under consideration. While these lightcurves may be fainter compared to those observed in the GW170817 event, they may be detectable through targeted searches using observatories such as Rubin Observatory~\citep{LSST:2008ijt}. Lastly, in 
Sec.~\ref{sec:conclusion}, we conclude our findings.

\begin{figure*}[htb]
    \centering
    \includegraphics[width=\textwidth]{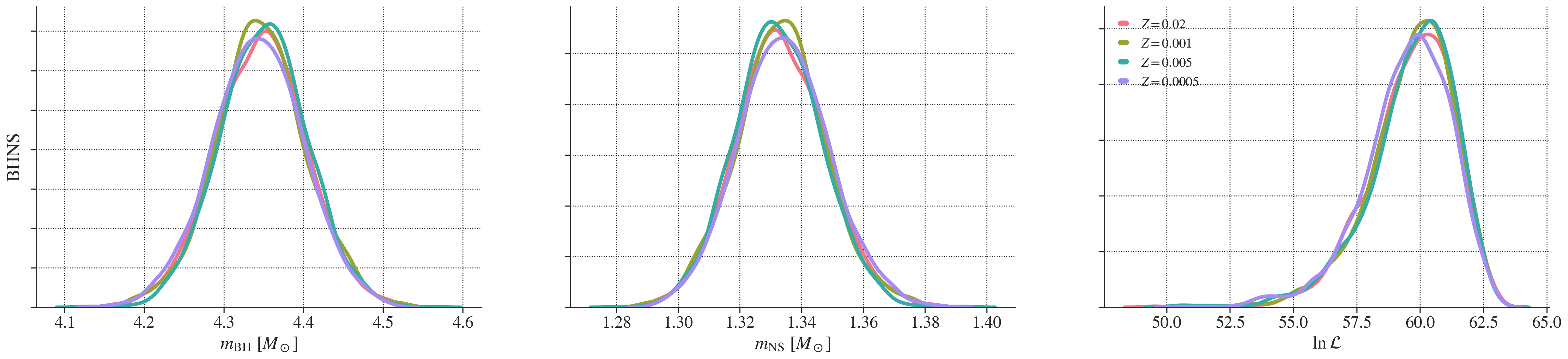}
    \includegraphics[width=\textwidth]{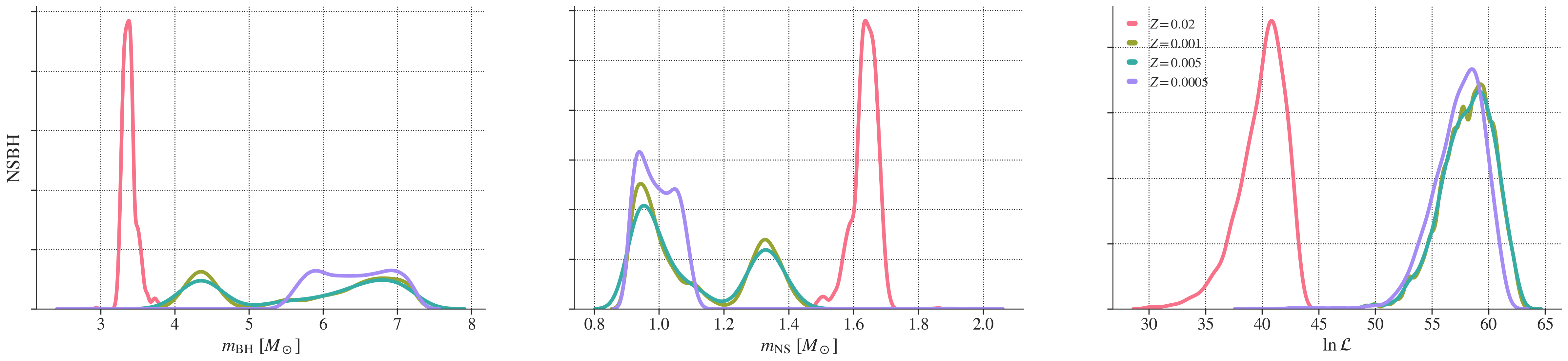}
    \caption{Posterior densities for the mass of GW230529 as inferred using population-inspired \ac{BHNS} (top panel) and \ac{NSBH} (bottom panel) priors described in Sec.~\ref{sec:bhns-nsbh}. In the right-hand column, we include the log-likelihood ratio distribution.}
    \label{fig:bhns-posterior-metallicity}
\end{figure*}

\section{Methods}\label{sec:methods}

We infer GW230529's properties by analysing 128s of LIGO Livingston data using the Bayesian \ac{pe} library \BILBY~\citep{Ashton:2018jfp} and the posterior sampling algorithm \DYNESTY~\citep{Speagle:2019ivv}. We assume a noise power spectrum given by the median estimate provided by \BAYESLINE~\citep{Littenberg:2014oda} and use frequencies in the range of $20-1792$ Hz for evaluating the \ac{GW} Transient log-likelihood ratio $\ln \mathcal{L}$. Furthermore, to speed up the
likelihood evaluation, we include heterodyning (also known as relative binning; \citep{Cornish:2010kf, Zackay:2018qdy, Cornish:2021lje, Krishna:2023bug}). 
For our analysis, we use the quasi-circular frequency-domain phenomenological waveform model, \texttt{IMRPhenomNSBH}~\citep{Foucart:2013psa,Foucart:2014nda,Khan:2015jqa,Dietrich:2019kaq,Thompson:2020nei}. This waveform approximant models signals using the dominant (quadrupole) harmonic and is specifically designed to model \acp{GW} emanating from \acl{BHNS} mergers with mass ratios ranging from equal-mass up to $q=m_\mathrm{BH}/m_\mathrm{NS}=15$. It also incorporates \ac{BH} spins up to a dimensionless value of $\chi_\mathrm{BH}=0.5$ and includes matter-effects through tidal parameters $\Lambda_\mathrm{NS}$ ranging from 0 to 5000. We assess the impact of waveform systematics in App.~\ref{app:wfsys}, finding them to be negligible for the purposes of our study.

However, unlike the original analyses, we exclude the marginalization over the systematic error in the measured astrophysical strain due to the detector calibration. This error is sub-dominant to the systematic errors from waveform modelling and prior choices, and we, therefore, neglect it~\citep{Ashton:2021cub}. Furthermore, as discussed in Sec.~\ref{sec:priors}, we apply astrophysically motivated mass and spin priors while using uninformative priors for all other parameters. It is important to note that, for all our analysis, we utilise the \ac{BH} and \ac{NS} masses viz $m_\mathrm{BH}$ and $m_\mathrm{NS}$ priors derived from population synthesis models as two distinct, one-dimensional independent priors.

While various assumptions may be made to model the formation pathway of this system, one needs to determine the relative probability of two models (in this case, binary formation process) given the data. The Bayesian evidence, $\mathcal{Z}$, quantifies this support. Varying prior assumptions can yield differing parameter estimates; therefore, the Bayes Factor, $\mathrm{BF}^{A}_{B}=\mathcal{Z}_A/\mathcal{Z}_B$, indicates whether the prior assumption $A$ is favoured or disfavoured compared to $B$ based on the data. This comparison is particularly crucial as strong prior assumptions may bias the posteriors towards potentially arbitrary values at the expense of the evidence. Furthermore, we also compare the log-likelihood ratio distribution since certain models allow for broader priors and incur a higher Ockham penalty.

\section{Astrophysical Implications}

\citet{LVK:2024elc} found no conclusive evidence either supporting or refuting the presence of tidal effects in the GW230529 signal. This makes it difficult to determine the nature of the compact objects involved. However, they showed that the (lighter) secondary component appears to be a \ac{NS}, while the (heavier) primary is likely a \ac{BH}, when using observationally constrained priors. In this section, we investigate which of the binary components formed first and determine their formation mechanism.

\subsection{BHNS or NSBH?}\label{sec:bhns-nsbh}

Black hole-neutron star binaries can generally be divided into two categories: (1) \ac{BHNS} mergers, where the \ac{BH} forms first, and (2) \ac{NSBH} mergers, in which the \ac{NS} forms first. While \acp{BHNS} are the dominant binaries according to population synthesis studies~\citep{Chattopadhyay:2020lff}, \acp{NSBH} are more exciting as they can form radio pulsars, generate \ac{KN} \citep{Barbieri:2019kli}, and lead to precise measurement of \ac{NS} spins \citep{Gupta:2024bqn}. 
To discriminate between the two, we analyze GW230529 using different predicted distributions of the detectable \ac{BHNS} and \ac{NSBH} masses and spins for different metallicity choices ($Z=0.02,0.001,0.005,0.0005$) from the base model of~\citet{Chattopadhyay:2022}.
These models are expected to be representative of the population of such binaries obtained from \ac{GW} observations. 

As can be seen in Fig~\ref{fig:bhns-prior-metallicity}, the \ac{BHNS} (top panel) and \ac{NSBH} (bottom panel) systems exhibit distinct mass spectra, with \acp{BH} in \ac{BHNS} systems being more massive. This is because the heavier star evolves faster and remains massive enough to form a \ac{BH} even after the mass transfer \citep{Hurley:2000, Belczynski:2010}. In contrast, \ac{NSBH} systems have more mass-symmetric progenitors, more so at higher metallicities. The heavier star transfers enough mass to its companion to form a \ac{NS}, whereas its companion becomes a \ac{BH}. 
However, at lower metallicities, reduced stellar winds lead to reduced mass loss, creating larger \acp{BH} that merge in a shorter time scale~\citep{Hurley:2000}. 

We use \ac{BH} spins derived from fits in~\citet{Chattopadhyay:2020lff} for \ac{NSBH} mergers but restricted to $\chi_\mathrm{BH}=0.5$ as \texttt{IMRPhenomNSBH} is not calibrated for $\chi_\mathrm{BH}>0.5$. Moreover, attaining spins greater than this would require an unphysical amount of matter accretion~\citep{Mandel:2020lhv}. Although tidal synchronization can lead to higher \ac{BH} spins, the fraction of such binaries is small~\citep{Chattopadhyay:2020lff}. For the \ac{BHNS} case, we set $\chi_\mathrm{BH}=\chi_\mathrm{NS}=0$  as the rotational velocities of \acl{NS} are anticipated to diminish over time due to electromagnetic radiation~\citep{Fragos:2014cva, Ma:2019cpr}. As ground-based \ac{GW} observatories are only expected to detect these binaries close to the merger, we assume that the \ac{BH} and \ac{NS} objects have negligible spin. Thus, for the second analysis, we effectively assume that the \textit{source of GW230529 is a non-spinning \ac{BHNS} binary}.

Figure~\ref{fig:bhns-posterior-metallicity} shows an obvious trend; the log-likelihood ratio distribution associated with the \ac{BHNS} (top panel) has a higher median log-likelihood ratio value and contains a prominent peak compared to the \ac{NSBH} case (bottom panel) for the same metallicity choices. Further, the \ac{BHNS} hypothesis is preferred with a $\ln \mathrm{BF}^\mathrm{BHNS}_\mathrm{NSBH} > 17(9)$ for metallicity choice of $Z=0.02 (0.0005)$. Therefore, we will assume that GW230529 is a \ac{BHNS} merger for the remainder of the article. Note that our results are robust against waveform systematics (See App.~\ref{app:wfsys}).

\begin{figure*}[htb]
    \centering
    \includegraphics[width=0.98\textwidth]{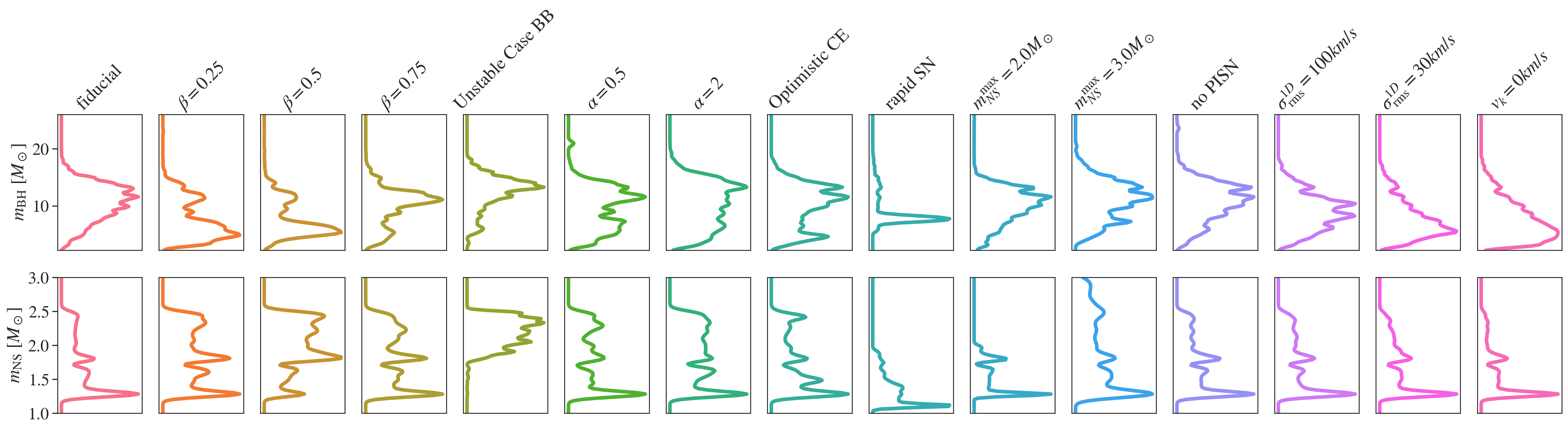}
    \caption{Predicted \ac{BHNS} component mass distributions. Each row shows the 15 different population synthesis variations used by~\citet{Broekgaarden:2021iew}.}
    \label{fig:prior}
\end{figure*}

\subsection{Constraints on binary evolution}\label{sec:priors}

The pathway leading to \ac{BHNS} mergers is still debated. The prevailing hypothesis suggests that these mergers arise from two massive stars that were born in a binary and evolved in isolation, typically involving the \ac{CE} episode that tightens the binary's orbit~\citep{Mandel:2021smh}. However, accurately estimating the rates of these mergers is challenging for several reasons.

Firstly, the physical processes that govern the evolution of massive binary star systems, including the dynamics of the \ac{CE} phase \citep{Ivanova2013}, mass transfer efficiency between binary components, and the kicks imparted to stars during \acf{SNe}, are complex and poorly understood, leading to considerable uncertainty \citep{Belczynski2022}. Secondly, uncertainties arise due to the star formation rate and the metallicity distribution within star-forming gas across cosmic time \citep{Langeroodi2023, Garcia2024arXiv}. Together, they significantly impact the detectable \ac{BHNS} mass distributions.

\begin{figure*}[htb]
    \centering
    \includegraphics[width=\textwidth]{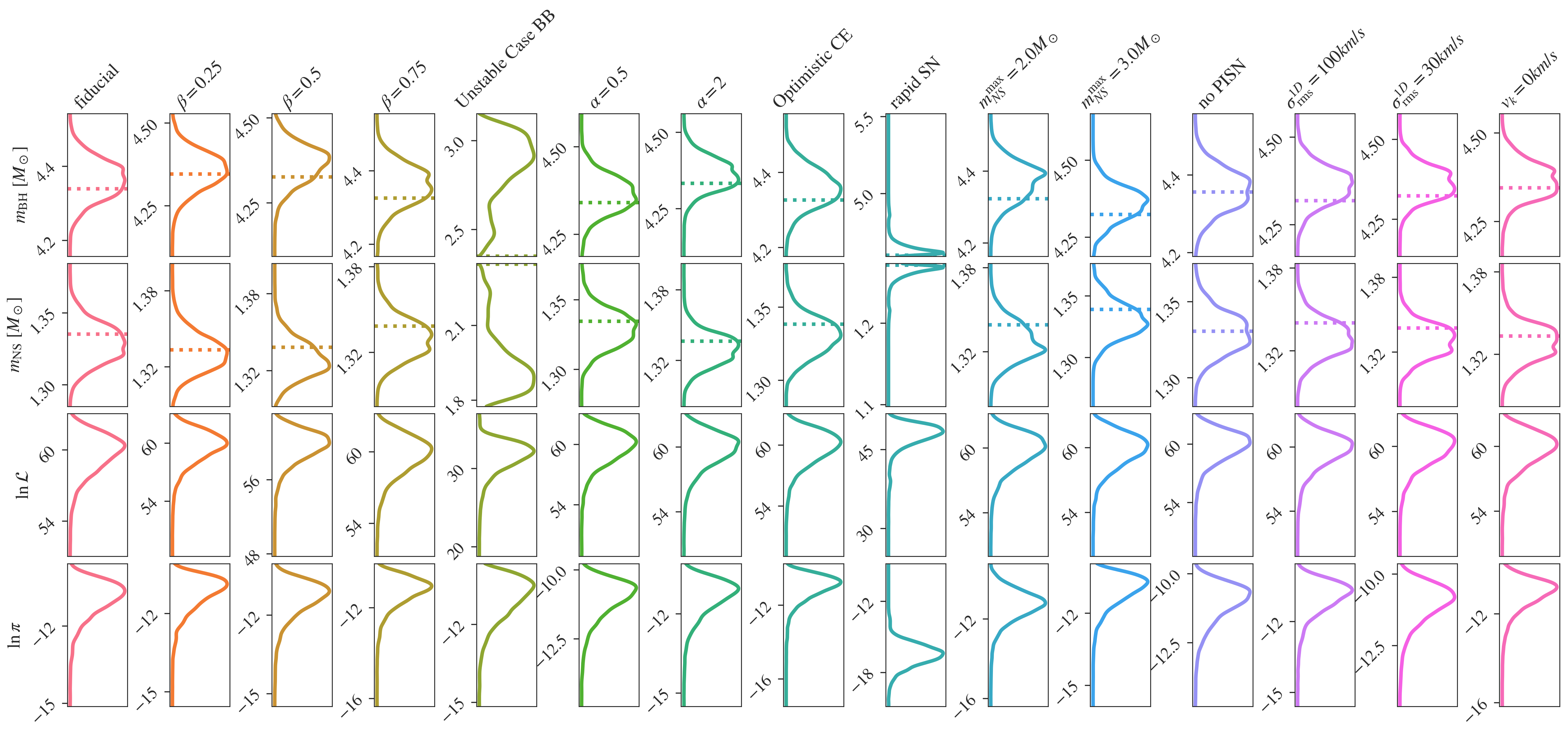}
    \caption{Posterior densities for the mass of GW230529 under different binary evolution assumptions. We also include the distributions of the log-likelihood and log-prior of the posterior samples. The dotted lines in the first two rows denote the maximum likelihood estimates of $m_\mathrm{BH}$ and $m_\mathrm{NS}$ respectively.}
    \label{fig:posterior}
\end{figure*}

Since we are interested in determining the physical process leading to GW23059's formation, we only focus on the uncertainties related to the physical processes. To that end, we, following \citet{Broekgaarden:2021iew}, assumed 15 different binary population synthesis predictions for \ac{BHNS} mass distribution. These models implement variations to the \textit{fiducial} model in different aspects of physics relevant to the binary evolution, such as mass transfer efficiency between binary components and the kicks imparted to the stars. We briefly summarise them below.

The $\beta=0.25, 0.5, 0.75$ models assume fixed mass transfer efficiencies. These models represent the fraction of mass lost by the donor star that its companion accretes. On the other hand, the ``unstable case BB'' model involves an unstable mass transfer phase from a stripped post-helium-burning star onto a \ac{BH}.

For the $\alpha=0.5$ and $\alpha=2$ models, a pessimistic \ac{CE} scenario is assumed, where the donor stars struggle to successfully eject their envelopes, resulting in efficiency parameters of 0.5 and 2, respectively. Conversely, the ``optimistic'' \ac{CE} scenario posits that these systems can survive such challenges.

To avoid creating a remnant mass gap between \acp{NS} and \acp{BH}, which contradict observations from X-ray binaries, \citet{Broekgaarden:2021iew} use a delayed remnant mass prescription in their simulations. However, the ``rapid \ac{SNe}'' model adopts a faster remnant mass prescription and is consistent with current observations.

The models labeled as $m_{\rm{NS}}^\mathrm{max} = 2M_\odot$ and $m_{\rm{NS}}^\mathrm{max} = 3M_\odot$ set the maximum mass of \ac{NS} to 2 and $3M_\odot$, respectively. It's worth noting that the latter case may be considered unrealistic since the maximum mass supported by current \ac{EoS} for non-rotating \ac{NS} is $\lesssim 2.9M_\odot$~\citep{Godzieba:2020tjn}.

\begin{figure*}[htb]
    \centering
    \includegraphics[width=\textwidth]{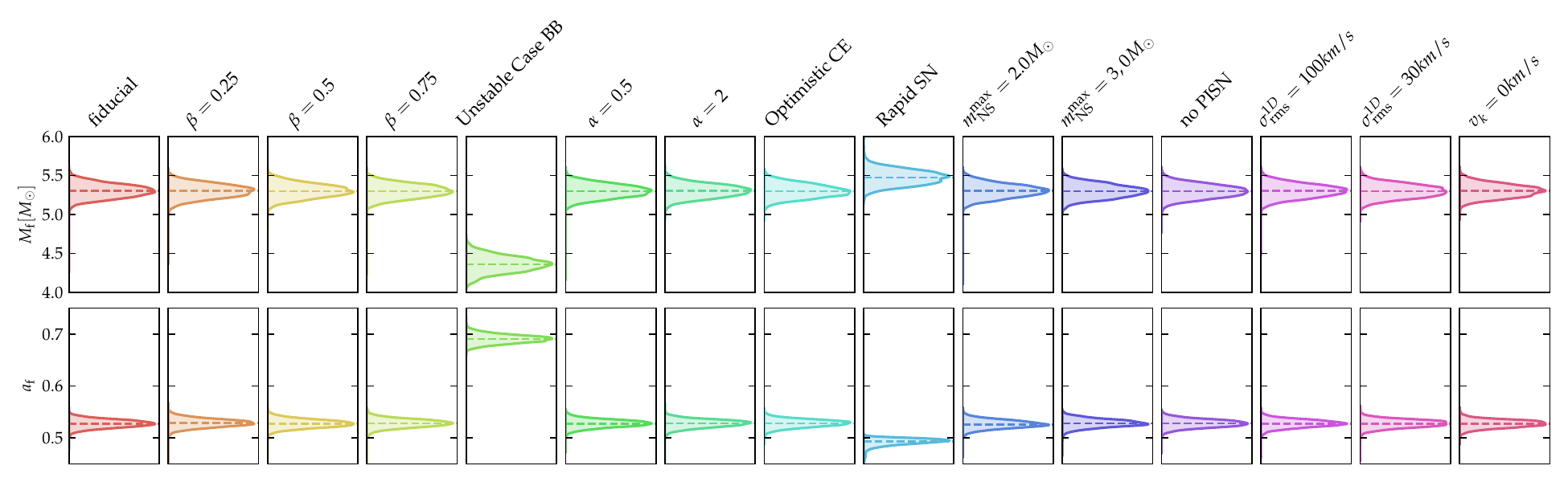}
    \caption{Posterior densities for the remnant \ac{BH} mass and spin under different binary evolution assumptions. These quantities are computed from \ac{nr} fits of \ac{BHNS} binaries, using the posteriors from Sec.~\ref{sec:priors} and the \ac{EoS} information coming from
    multi-messenger analyses of GW170817 and AT2017gfo.
    For most of the scenarios considered we find $M_f = 5.3^{+0.1}_{-0.1} M_{\odot}$ and $a_f = 0.53^{+0.01}_{-0.01}$. }
    \label{fig:remnant}
\end{figure*}

The ``no PISN" prescription excludes the pair-instability process, responsible for the scarcity of first-generation \acp{BH} with masses between $65$--$120M_\odot$. Additionally, the $\sigma_{\rm{cc}} = 30$ km/s and $\sigma_{\rm{cc}} = 100$ km/s prescriptions explore variations of natal kicks compared to the one-dimensional root-mean-square velocity dispersion of $\sigma_{\rm{cc}} = 265$ km/s used in the fiducial model. These lower values can occur in ultra-stripped \ac{SNe} and electron-capture \ac{SNe}, leading to reduced binary disruption. Finally, the $v_{\rm{k,BH}} = 0$ km/s assumption posits that \acp{BH} receive no supernova natal kicks.

Fig~\ref{fig:posterior} summarises our findings when assuming these populations. We observe that, except for the ``unstable case BB" and ``rapid \ac{SNe}" models, the inferred posteriors are largely in agreement with each other. This alignment was anticipated, given that these two population models restrict the \ac{BH} mass to a range outside the support for this signal. Moreover, we do not observe any trends in the log-prior distribution for the other models, which indicates that all the information about the evidence for and against binary evolution is contained in the distribution of the log-likelihood ratio. But we also do not observe any differences in the shape and location of the log-likelihood ratio distribution, indicating that all models are equally likely. This is not surprising since the signal's \ac{SNR} is low. Finally, it is noteworthy that, for the majority of our analysis, we determine the \ac{BH} and \ac{NS} masses to be $m_\mathrm{BH}=4.3^{+0.08}_{-0.09}M_\odot$ and $m_\mathrm{NS}=1.4^{+0.02}_{-0.02}M_\odot$, respectively, while finding that the $\ln \mathrm{BF}^\mathrm{fiducial}_\mathrm{pop-agnostic} \sim 1$.


\section{Multimessenger Predictions}

\subsection{Remnant properties}\label{sec:remnant-mma}

\begin{figure*}[htb]
    \centering
    \includegraphics[width=\textwidth]{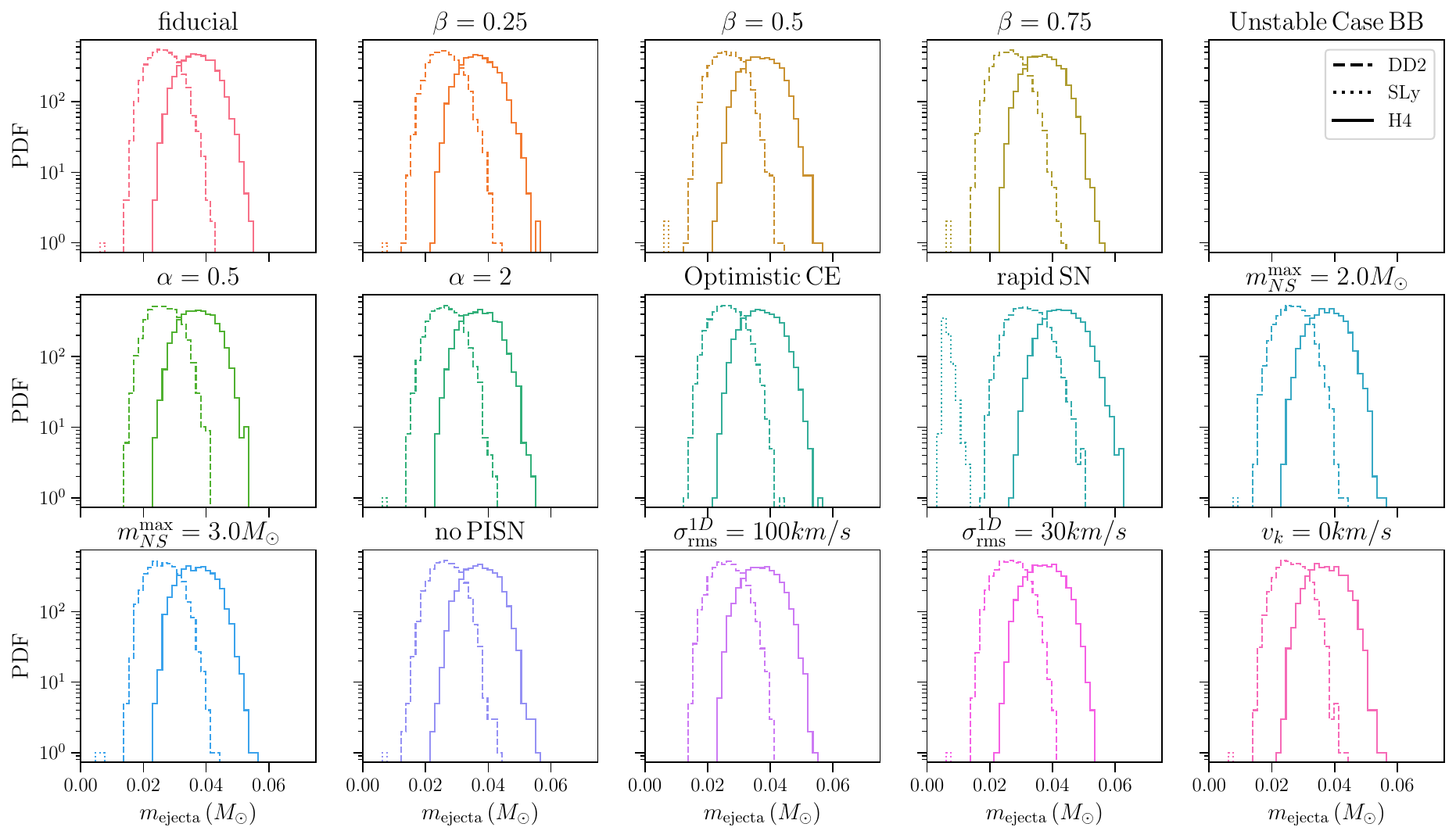}
    \caption{The probability distribution function (PDF) for the total ejecta mass for different EoS and corresponding to the various population models. No ejecta was obtained for the \texttt{APR4} EoS.}
    \label{fig:ejecta}
\end{figure*}

\begin{figure*}[htb]
    \centering
    \includegraphics[width=\textwidth]{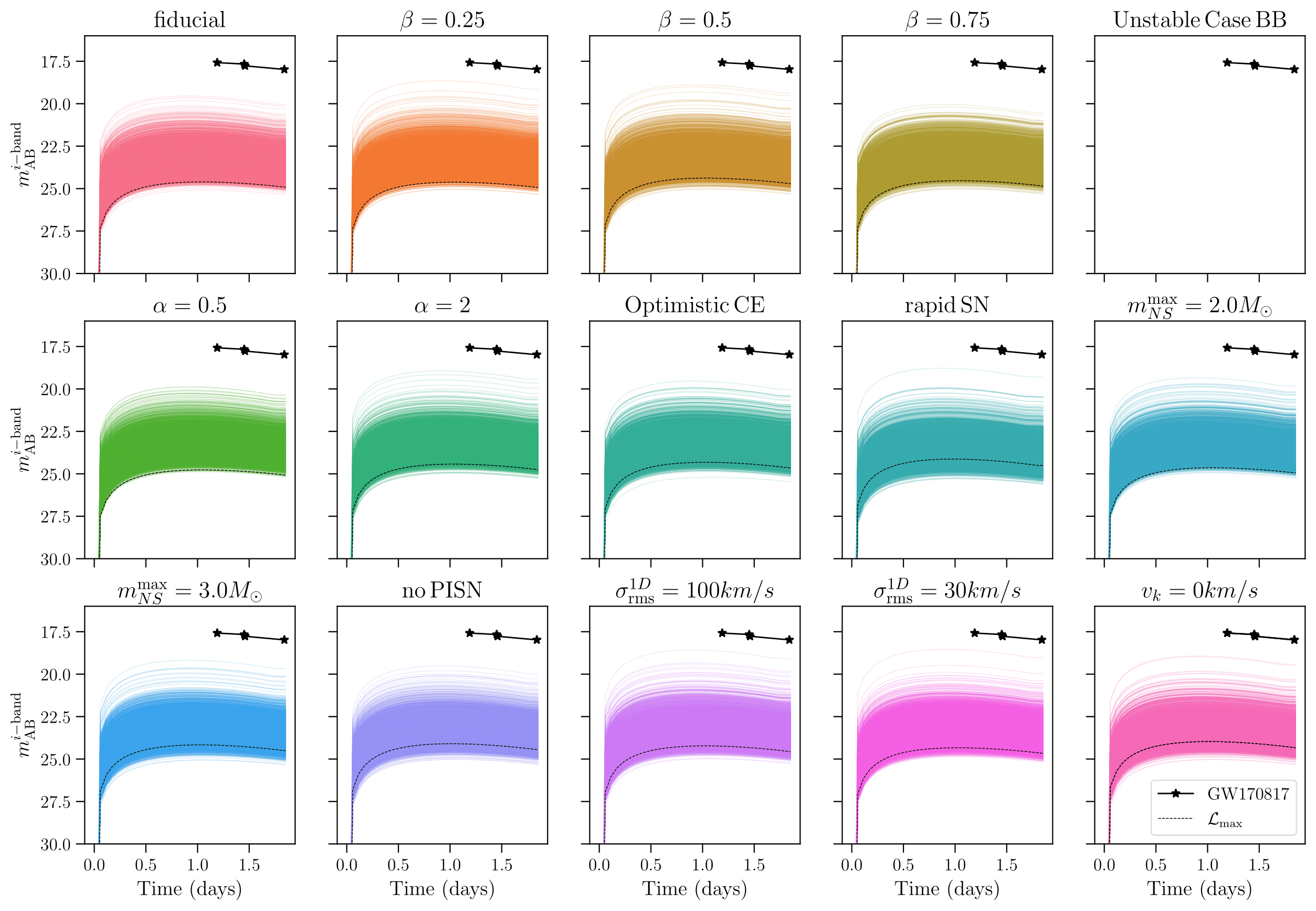}
    \caption{\acp{KN} light curves in the $i-$band corresponding to the different population models for the \texttt{H4} EoS. The dotted line shows the light curve associated with the maximum likelihood parameter estimates for each model.}
    \label{fig:kilonovae}
\end{figure*}

Assuming GW230529 is a \ac{BHNS} merger, estimates of the final mass and spin of the remnant \ac{BH} can be made using the models of \citet{Zappa:2019ntl, Gonzalez:2022prs}. 
Rather than employing the uninformative posteriors on the tidal parameters obtained from our \ac{pe}, we combine samples obtained using the the \texttt{IMRPhenomNSBH} waveform with the \acp{EoS} of \citet{Breschi:2024qlc}. This set was obtained by performing joint analysis of GW170817~\citep{LIGOScientific:2017vwq, LIGOScientific:2018hze} and AT2017gfo~\citep{Villar:2017wcc}, including information from \ac{nr} in the \acp{KN} model employed and folding in pulsar mass and radii measurements~\citep{Miller:2021qha, Romani:2022jhd, Vinciguerra:2023qxq}. 
%

Figure~\ref{fig:remnant} shows the predicted remnant property distribution for various formation model assumptions.
We find that all models, with the exception of the ``Unstable case BB'' and ``the Rapid \ac{SNe}'', predict the formation of a $M_f = 5.3^{+0.1}_{-0.1} M_{\odot}$ \ac{BH} remnant, with a dimensionless spin $a_f = 0.53^{+0.01}_{-0.01}$. 

Applying the classification proposed in Sec.~II of \citet{Gonzalez:2022prs}, we further find
high probability ($\geq 99\%$) of the \ac{NS} being at least partially tidally disrupted during the coalescence, with about $13\%$ probability of such disruption happening far from the system's innermost circular orbit. This value is consistent with the findings of \cite{LVK:2024elc}, indicating that one would expect to observe a suppressed but nonetheless present ringdown signal and a possible electromagnetic counterpart.

\subsection{Ejecta Mass} \label{sec:ejecta}

During a \ac{BHNS} merger, the \ac{NS} can get tidally disrupted. The neutron-rich ejecta from such disruption can emit \acp{KN}~\citep{Metzger:2016pju} provided the \ac{NS} disrupts before it enters the innermost circular orbit $(R_{\rm ISCO})$ corresponding to the \ac{BH}. If the \ac{NS} sheds mass after crossing this orbit, the \ac{BH} swallows the ejecta, producing no \ac{KN}. Hence, the distance at which the \ac{NS} starts shedding mass must be greater than $R_{\rm ISCO}$ to produce a transient. The ratio of the mass-shedding radius and $R_{\rm ISCO}$ increases with the \ac{BH} spin and decreases with the binary's mass ratio and the \ac{NS}'s compactness for aligned spin. For anti-aligned cases, the mass-shedding limit increases significantly, disfavoring tidal disruption. 
Thus, mass-symmetric binaries with a rapidly spinning \ac{BH} are conducive to generating \acp{KN}. Similarly, less compact \acp{NS} (coming from stiff \ac{EoS}) are easier to disrupt for aligned spin case, which is also favorable for \acp{KN} production.
We utilize the isotropic framework proposed in \citep{Arnett1980-os, Arnett1982-wf, Chatzopoulos2012-km, Villar:2017wcc, Kashyap:2019ypm, Gupta:2024} to calculate \ac{KN} light curves resulting from r-process nucleosynthesis in \ac{BHNS} mergers (see \citealt{Kawaguchi:2024hdk} for comparison and angle-dependence on light curve). The \acp{KN} lightcurves are characterized by the ejecta mass, velocity and opacity, which are informed by and related to the binary parameters using up-to-date \ac{nr} simulations~\citep{Kyutoku2018-yq, Barbieri2019-ar, Lattimer1976-zs, Lattimer1974-tz, Kyutoku2021-mh, Kruger2020-bw}. We use the posterior samples obtained in the Sec.~\ref{sec:priors} and calculate the mass of total ejecta, dynamical ejecta, and unbound disk ejecta using the fits from~\citet{Kruger:2020gig} and \citet{Raaijmakers:2021slr}.

This calculation is performed using four different \acp{EoS}: \texttt{APR4}~\citep{Akmal:1998cf}, \texttt{SLy} ~\citep{Chabanat:1997un, Douchin:2001sv}, \texttt{DD2}~\citep{Typel:2009sy, Hempel:2009mc} and \texttt{H4} \citep{Glendenning:1991es, Lackey:2005tk, Read:2008iy} (see Fig.~\ref{fig:eos} in Appendix~\ref{app:KNe} for corresponding mass-radius curves).
%
%
Among these, \texttt{APR4} and \texttt{SLy} give relatively more compact NSs, whereas \texttt{DD2} and \texttt{H4} result in less compact ones. Thus, the latter two will produce more ejecta than the former. We find that \texttt{APR4} does not give any ejecta for samples corresponding to any population models. Fig.~\ref{fig:ejecta} shows the probability distribution of total ejecta for the remaining three \ac{EoS}. Across population models, the total ejecta mass corresponding to \texttt{SLy} $\lesssim 0.01M_{\odot}$, \texttt{DD2} $\lesssim 0.05M_{\odot}$, and \texttt{H4} $\lesssim 0.07M_{\odot}$. Most ejecta are produced for the samples corresponding to the ``rapid \ac{SNe}" population model, which is due to its preference for a more symmetric binary (c.f. Fig.~\ref{fig:posterior}). On the other hand, the ``Unstable Case BB" population model results in no ejecta, due to its preference for more massive (hence, more compact) \ac{NS}. Note that both ``rapid \ac{SNe}" and ``Unstable Case BB" population models are disfavored compared to others, as discussed in Sec.~\ref{sec:priors}.

\subsection{Kilonova Lightcurves} \label{sec:lightcurves}

Choosing the most optimistic scenarios, namely those corresponding to \texttt{H4} \ac{EoS}, we calculate the bolometric \acp{KN} light curves (see Appendix~\ref{app:KNe} for \texttt{DD2}). We convert them to bandwise light curves for the \textit{ugrizy} filters (see \citet{Gupta:2023evt} and Appendix~\ref{app:KNe} for more details). Figure~\ref{fig:kilonovae} shows $i$-band curves together with the curve for maximum likelihood $(\mathcal{L}_{\rm max})$ binary parameters. We also compare the points from the detected light curve of GW170817 \citep{Guillochon_2017, Villar:2017wcc} in the $i$-band.

From Fig.~\ref{fig:kilonovae}, we note that only the “Unstable case BB” mass transfer population model does not support a \ac{KN} after the merger, a population model disfavored by our analysis.
For all other population models, the \ac{KN} is expected to be considerably dimmer than the one observed for GW170817.  This is expected, given that the \ac{KN} flux is inversely proportional to the distance squared and that GW230529 is located approximately five times further than GW170817 \citep{LVK:2024elc}. For all population models, the curves corresponding to the $\mathcal{L}_{\rm max}$ binary parameters prefer comparatively dimmer \acp{KN}, which is because the $\mathcal{L}_{\rm max}$ parameters correspond to more mass-asymmetric binary located further away.

\begin{table}[htbp] 
\centering
\caption{\label{tab:kne_maxlum} The (optical) bandwise peak luminosities and decay in luminosity one day after peak for the \acp{KN} with $\mathcal{L}_{\rm max}$ binary parameters corresponding to the ``fiducial" population model and the \texttt{H4} \ac{EoS}. For comparison, the single-exposure (30 s) bandwise limiting magnitudes corresponding to the Rubin Observatory $(m^{\rm lim}_{\rm RO})$ are also listed.}
\renewcommand{\arraystretch}{1}
\begin{tabular}{ c | c | c | c }
\hhline{----}
Band & $m^{\rm lim}_{\rm RO}$ ($m_{\rm AB}$) & Peak ($m_{\rm AB}$) & Decay ($m_{\rm AB}$) \\
\hhline{----}
$u$ & 23.9 & 26.63 & 1.85 \\
\hhline{----}
$g$ & 25.0 & 25.80 & 1.04 \\
\hhline{----}
$r$ & 24.7 & 25.05 & 0.59 \\
\hhline{----}
$i$ & 24.0 & 24.61 & 0.35 \\
\hhline{----}
$z$ & 23.3 & 24.27 & 0.16 \\
\hhline{----}
$y$ & 22.1 & 24.04 & 0.10 \\
\hhline{----}
\end{tabular}
\end{table}

Focusing on the \acp{KN} properties, we note that the luminosity in the $i$-band peaks a day after the merger for all population models except ``unstable case BB''. The peak luminosities corresponding to the $\mathcal{L}_{\rm max}$ binary parameters lie in the range $[23.97,24.67]$ mag. 
For the ``fiducial'' population model and $\mathcal{L}_{\rm max}$ binary parameters, Table~\ref{tab:kne_maxlum} shows the bandwise peak luminosities and its decay. The latter is the absolute difference between the peak luminosity and one day after it.

To gauge the \acp{KN} detectability, we compare the peak luminosity in each band with the corresponding limiting magnitude $(m^{\rm lim}_{\rm RO})$ of the Rubin Observatory for a single, 30s long exposure. While the \acp{KN} in the $y$-band is the brightest, Rubin wouldn't have observed it owing to its relatively poor sensitivity in this band. For the ``fiducial" population model, peak luminosities in the $gri$ bands come closest to the limiting magnitude threshold. However, a targeted observation with 600s exposure increases the limiting magnitude for the $g$ and $i$ bands to 26.62 and 25.62 \citep{Gupta:2022fwd, Branchesi:2023mws}, respectively, making the \acp{KN} visible in these bands. 

\section{Conclusions}\label{sec:conclusion}
In this work, we have studied the origins of GW230529, assuming that it is a binary formed via the classical isolated binary evolution via the \ac{CE} phase. As discussed in Sec~\ref{sec:priors}, by leveraging the \ac{BHNS} binary population synthesis model from \citet{Broekgaarden:2021iew}, we present compelling evidence that the system's properties are consistent with the predictions derived from the isolated binary evolution pathway of \ac{BHNS} systems. However, due to the event's relatively low \ac{SNR}, we face difficulties in identifying the underlying physical mechanism driving its formation unequivocally. However, we could rule out with confidence certain formation mechanisms such as the one involving the ``rapid \ac{SNe}'' or the ``unstable case BB''. 

Using our posterior samples and numerical fits from \citet{Zappa:2019ntl, Gonzalez:2022prs}, we infer the remnant's mass and spin, finding that it is consistent with a \ac{BH} with $M_f\sim 5.3M_\odot$ and $\chi_f \sim 0.53$. We also predict that there is a $\gtrsim 99\%$ probability that \ac{NS} is tidally disrupted during the merger and about 13\% probability that such disruption occurs outside $R_\mathrm{ISCO}$, leading to potential electromagnetic counterparts and a suppressed ringdown signal. 

We compute \acp{KN} light curves using our population-informed posterior samples and the isotropic framework proposed by~\citet{Arnett1980-os, Arnett1982-wf, Chatzopoulos2012-km, Villar:2017wcc}. We find that the i-band luminosity is dim enough not to be observed in a regular Rubin search but may be bright enough to be observed in a targeted one. 

Overall, GW230529 is a notable addition to the growing population of compact binaries observed in the \acl{GW} window. It marks the first of many more events with components with masses between $\sim 3$--$5M_\odot$, the so-called ``lower mass-gap'' that divides the \ac{NS} and \ac{BH} population.

\section*{Acknowledgements}
\label{sec:acknowledgements}
%
We thank Simon Stevenson and Aditya Vijaykumar for their comments and valuable suggestions. KC, IG, RK and BSS acknowledge the support through NSF grant numbers PHY-2207638, AST-2307147, PHY-2308886, and PHY-2309064.
RG acknowledges support from NSF Grant PHY-2020275
(Network for Neutrinos, Nuclear Astrophysics, and Symmetries (N3AS)). DC  is supported by the STFC grant ST/V005618/1. SB knowledge funding from the EU Horizon under ERC Consolidator Grant,
  no. InspiReM-101043372. DC and BSS thank the Aspen Center for Physics (ACP) summer workshop 2022 for setting up discussions that also contributed to this collaborative work. 
\texttt{TEOBResumS} is publicly available at \url{https://bitbucket.org/eob_ihes/teobresums}. The version
employed in this work is tagged via the arXiv submission number of the paper itself.
This research has made use of data, software and/or web tools obtained from the Gravitational Wave Open Science Center (https://www.gw-openscience.org),
a service of LIGO Laboratory, the LIGO Scientific Collaboration, the Virgo Collaboration, and KAGRA. This
material is based upon work supported by NSF’s LIGO Laboratory, which is a major facility fully funded by the
National Science Foundation. LIGO Laboratory and Advanced LIGO are funded by the United States National Science Foundation (NSF) as well as the Science and Technology Facilities Council (STFC) of the United Kingdom, the Max-Planck-Society (MPS), and the State of Niedersachsen/Germany for support of the construction of Advanced LIGO and construction and operation of the GEO600 detector. Additional support for Advanced LIGO was provided by the Australian Research Council. Virgo is funded through the European Gravitational Observatory (EGO), by the French Centre National de Recherche Scientifique (CNRS), the Italian Istituto Nazionale di Fisica Nucleare (INFN) and the Dutch Nikhef, with contributions by institutions from Belgium, Germany, Greece, Hungary, Ireland, Japan, Monaco, Poland, Portugal, Spain. KAGRA is supported by Ministry of Education, Culture, Sports, Science and Technology (MEXT), Japan Society for the Promotion of Science (JSPS) in Japan; National Research Foundation (NRF) and Ministry of Science and ICT (MSIT) in Korea; Academia Sinica (AS) and National Science and Technology Council (NSTC) in Taiwan. This document has been assigned the LIGO document number LIGO-P2400163.

\appendix
\

\section{Waveform systematics} \label{app:wfsys}

\begin{figure}
    \centering
    \includegraphics[width=0.45\textwidth]{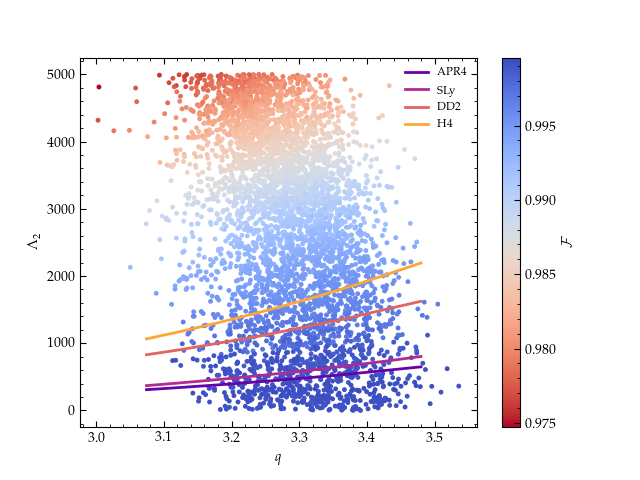}
    \includegraphics[width=0.45\textwidth]{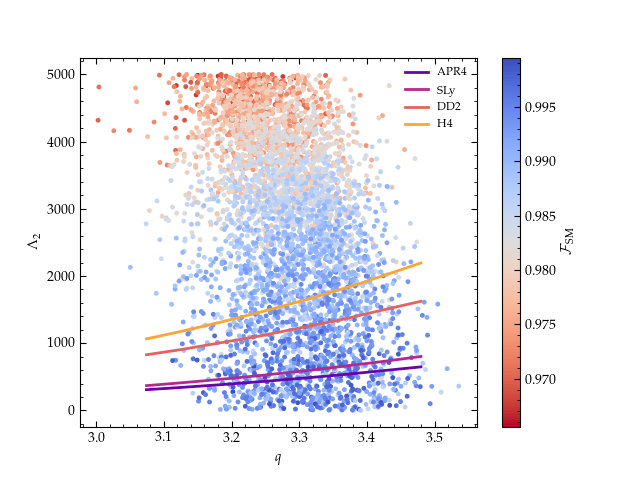}
    \caption{
    \label{fig:mismatches}
    Match $\mathcal{F}$ and sky-maximized match $\mathcal{F}_{\rm SM}$ between polarizations constructed using the dominant quadrupolar (top) and subdominant (bottom) modes of \texttt{TEOBResumS-GIOTTO} and \texttt{IMRPhenomNSBH}. 
    We consider posterior samples from the fiducial \ac{pe} and compute $\mathcal{F}$ between $20$ and $1796$ Hz. Depending on whether higher modes are employed or not, we find median mismatches of about $\sim 1.4\%$ ($0.8\%$) and maximum mismatches of $\sim3\%$ ($\sim2\%$), corresponding to systems with large tidal parameters $\Lambda_2 > 3000$.
    Overlaid, we also plot the tidal parameters obtained with a few selected \acp{EoS} assuming a fixed \ac{BH} source frame mass of $4.2 M_{\odot}$.}
\end{figure}

\citet{LVK:2024elc} demonstrated that while not dominant with respect to statistical uncertainties, systematic errors due to the choice of the employed \acl{GW} models are not negligible. Moreover, GW230529 inhabits a parameter space previously unexplored through observations. In this appendix, we elucidate that systematic uncertainties hold minimal significance within the framework of population-informed priors. Furthermore, we assert that the findings presented in this study remain robust regardless of the specific model chosen.

To substantiate this claim, we perform the following analyses: (i) we measure the mismatch between the two waveform models, namely the \texttt{IMRPhenomNSBH} and \texttt{TEOBResumS-GIOTTO} \citep{Akcay:2018yyh, Nagar:2019wds, Nagar:2020pcj, Gonzalez:2022prs}, the latter being an \acl{eob} model for compact objects informed by \acl{nr} simulations of merging \aclp{bbh} and \ac{BHNS} systems containing higher modes and tidal disruption~\citep{Buonanno:1998gg}. This study uses posterior samples obtained when assuming the fiducial model. (ii) We assess the robustness of mass parameter measurements against modelling systematics when employing the aforementioned waveform models.

\subsection{Comparison between \texttt{TEOBResumS-GIOTTO} and \texttt{IMRPhenomNSBH}} 

\begin{figure}[htb]
    \centering
    \includegraphics[width=0.33\textwidth]{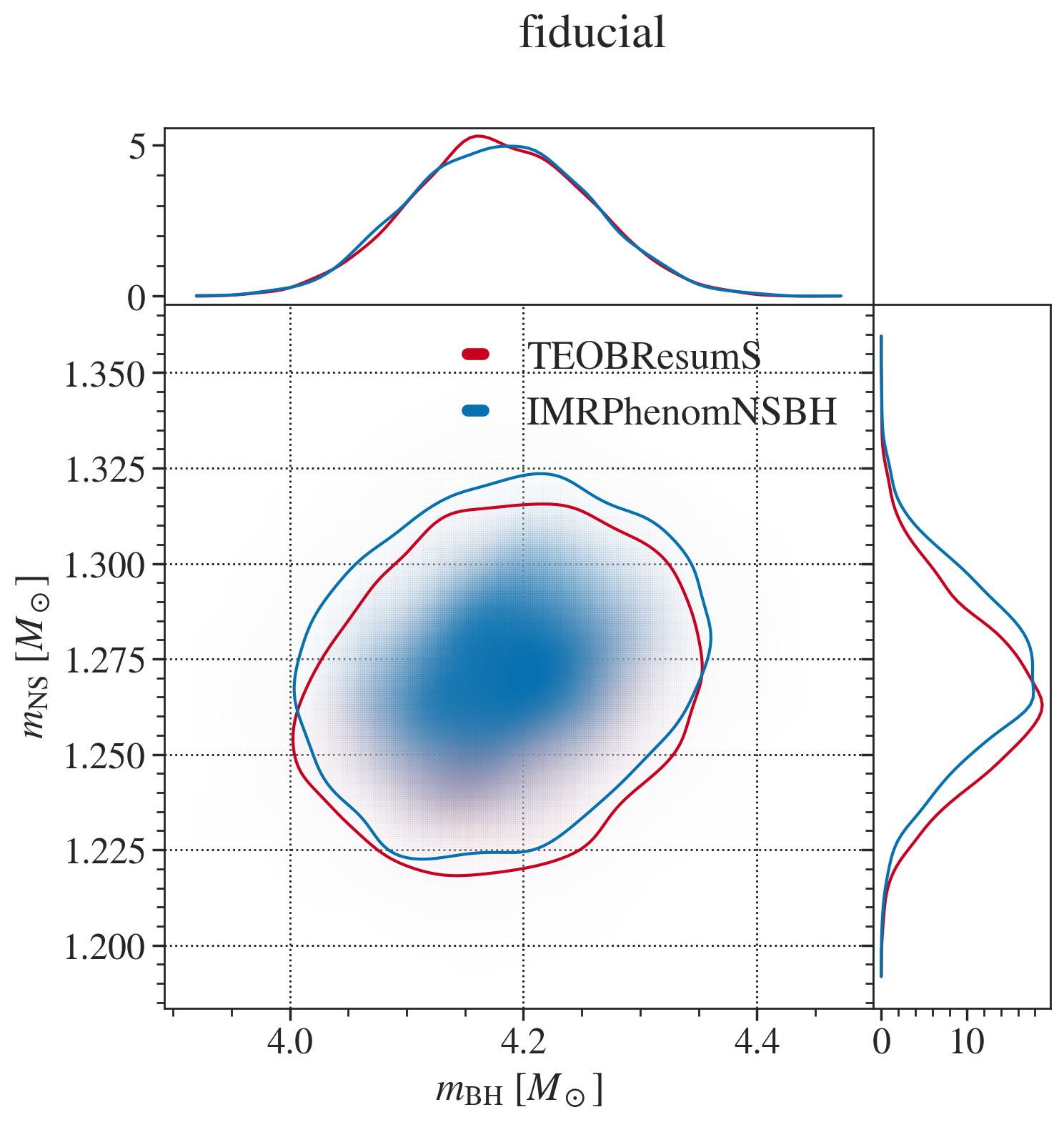}\\
    \includegraphics[width=0.33\textwidth]{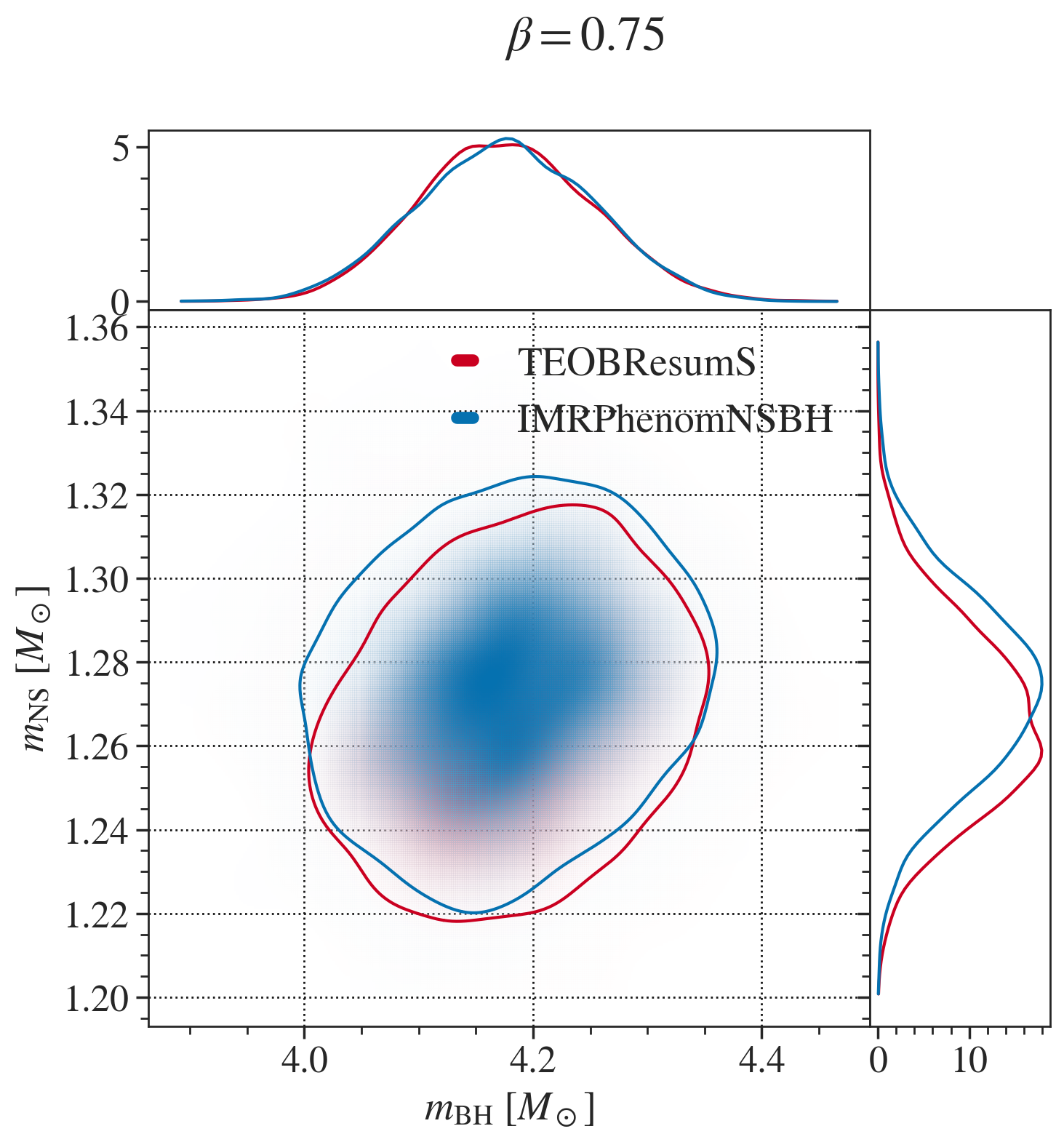}\\
    \includegraphics[width=0.33\textwidth]{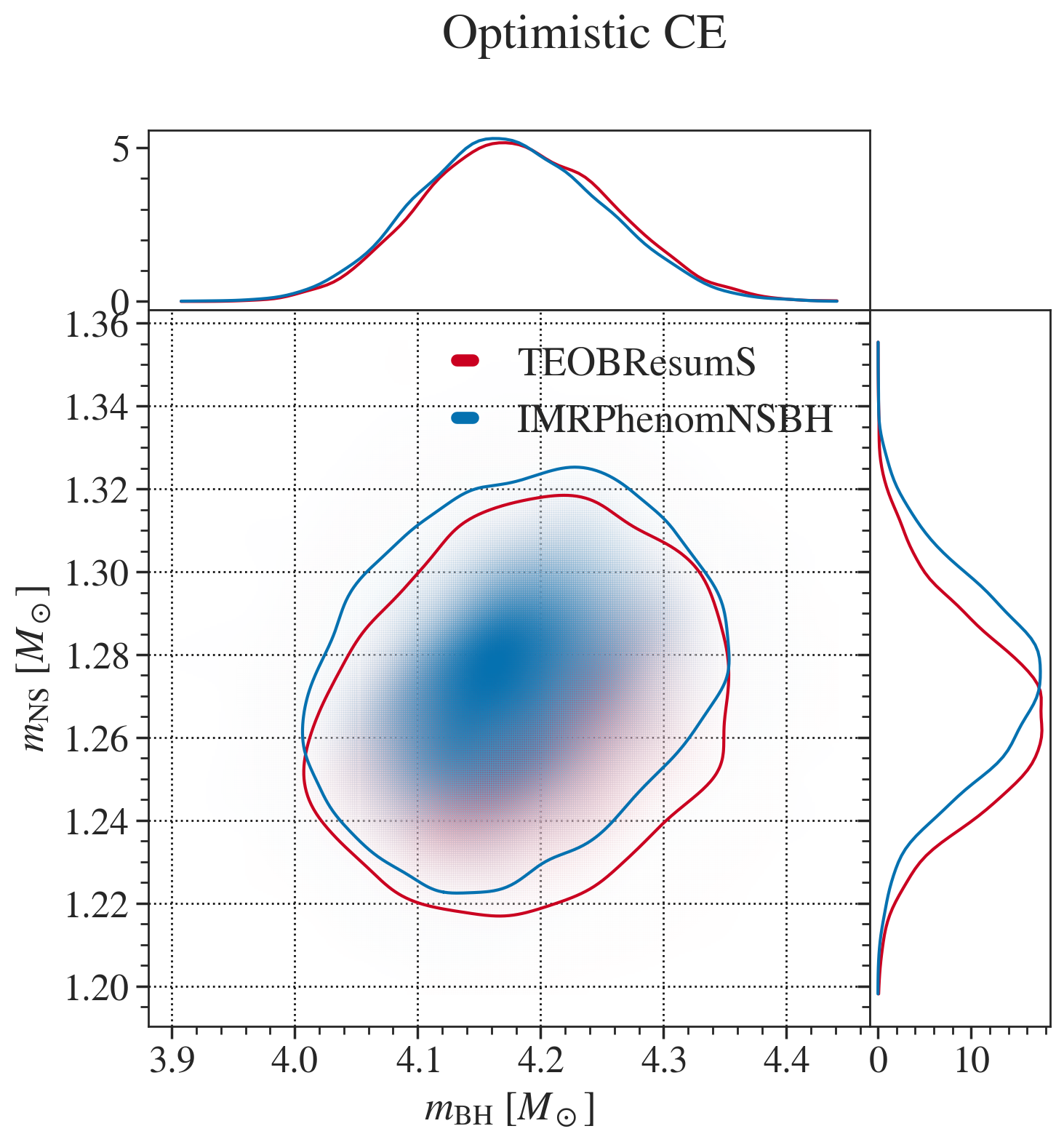}
    \caption{The posterior distribution of the primary and secondary source masses for two different waveform models --- \texttt{TEOBResumS-GIOTTO} and \texttt{IMRPhenomNSBH}. Each contour shows the 90\% credible intervals. We find that the resulting parameter estimates are robust to possible waveform systematics, with good agreement between the two waveform models.}
    \label{fig:teob-vs-imr}
\end{figure}

We quantify the discrepancy between the waveform models in terms of the match (or faithfulness)~\citep{Cutler:1994ys, Apostolatos:1995pj}:
\begin{equation}\label{eq:A1}
    \mathcal{F} = \max_{t_{\rm ref}, \phi_{\rm ref}} \frac{(h|k)}{\sqrt{(h|h)(k|k)}} \, ,
\end{equation}
where $t_{\rm ref}$ and $\phi_{\rm ref}$ are a reference time and phase, $h$ and $k$ are the two waveforms considered and
\begin{equation}
    (h|k) = 4 \Re\int_{f_{\rm min}}^{f_{\rm max}} \frac{\tilde{h}(f)\tilde{k}^*(f)}{S_n(f)} df \, .
\end{equation}
Since \texttt{TEOBResumS-GIOTTO} contains higher-order radiation multipoles\footnote{In detail, we employ the $(\ell, |m|) = (2,1), (2,2), (3,3), (4,4)$ modes.}, the definition above is not independent of the extrinsic parameters of the binary. We, therefore, maximise over the sky position and polarization by computing the sky-maximized match $\mathcal{F}_{\rm SM}$ as defined in~\cite{Harry:2017weg, Chandra:2022ixv}. This statistic reduces to the Eq.~\eqref{eq:A1} when only $(2, |2|)$ modes are employed. We then compute this quantity over the frequency interval $[20, 1796]$ Hz and employ the \acl{PSD} $S_n$ of GW230529. Crucially, to avoid noise artefacts in the fast Fourier transform, we generate the waveforms from $18$ Hz and taper them at the beginning.

We find that the mismatches always lie below the $2.5\%$ threshold when only $(2,|2|)$ modes are considered and below $3.5\%$ when subdominant multipoles are included in the polarizations computation (see Fig.~\ref{fig:mismatches}).
As expected, both $\mathcal{F}$ and $\mathcal{F}_{\rm SM}$ decrease with increasing tidal parameters:
if one reduces the interval of $\Lambda_2$ considered to $[0, 3000]$, the maximum values for the figures above shift to $1.2\%$ and $2.3\%$ respectively.
This is because neutron stars characterized by large values of $\Lambda_2$ are more easily disrupted, leading to growing differences in the model's predictions in a region poorly explored by numerical relativity simulations.

To understand whether the matches obtained are ``acceptable", they should, in principle, be compared to some accuracy requirement, i.e. to some theoretical threshold below which one may expect waveform systematics to appear~\citep{Cutler:1994ys, Lindblom:2008cm, Damour:2010zb, Chatziioannou:2017tdw, Toubiana:2024car}. While correctly assessing accuracy requirements is largely a back-of-the-envelope estimate can be obtained employing the following threshold $\mathcal{F}_{\rm thrs}$~\citep{Damour:2010zb, Chatziioannou:2017tdw}:
\begin{equation}
    \mathcal{F}_{\rm thrs} = 1 - \frac{\epsilon}{2 \rho^2} \, ,
\end{equation}
where $\epsilon $ is the number of intrinsic parameters of the system. This choice leads to $\mathcal{F}_{\rm thrs} = 0.988$ ($\epsilon = 3$), 
smaller than $\sim65\%$ of the matches computed when only $(2,|2|)$ modes are
considered. When subdominant modes are further included in the analysis, inclination should be treated as an intrinsic source parameter and $\mathcal{F}_{\rm thrs} = 0.984$, smaller than $58\%$ of the mismatches computed.
As such, though some differences between models are certainly present, one should not expect large biases to appear.
 
\subsection{Re-analysis with \texttt{TEOBResumS}} \label{app:teob}

We repeated some of the population-informed parameter estimations using the time-domain waveform model \texttt{TEOBResumS-GIOTTO}, as shown in Fig~\ref{fig:teob-vs-imr}. The results indicate that the GW23059's source mass parameters are robust against modelling systematics.
This is consistent with the faithfulness study performed in the previous section. Although the models employ entirely independent descriptions of both point-mass and tidal sectors, differences are largely irrelevant: the high-frequency component of the data is dominated by noise and, as such, does not affect the inference.
Additionally, our prior constrains the spins of the system components to be effectively zero, further minimizing differences between models.

\section{Kilonova Lightcurves}\label{app:KNe}

\begin{figure}[htb]
    \centering
    \includegraphics[width=0.4\textwidth]{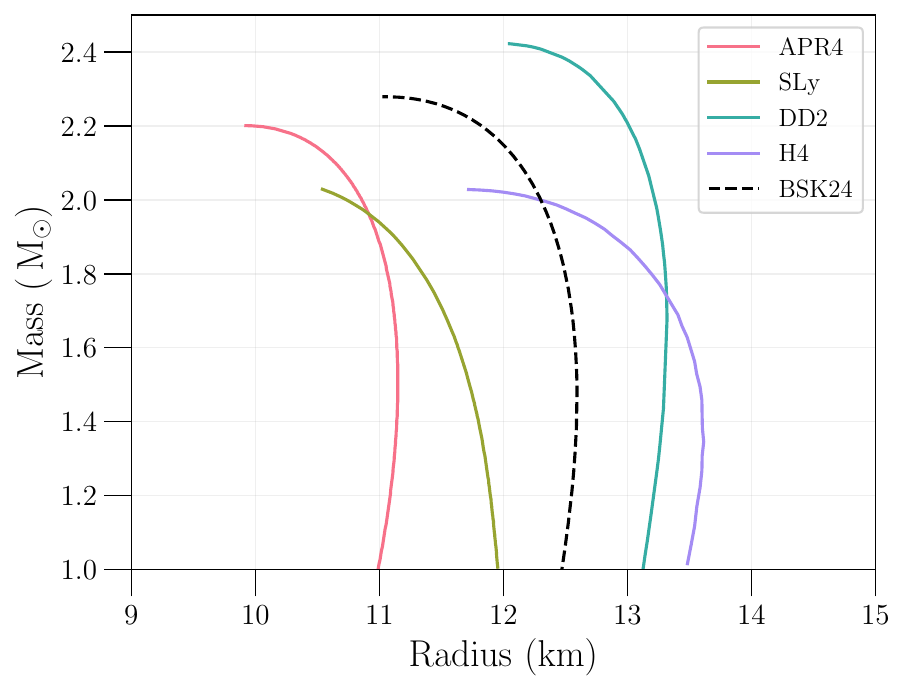}
    \caption{The mass-radius curves for the EoS chosen in this study. We also show the \texttt{BSK24} EoS which was used by \citet{LVK:2024elc}.}
    \label{fig:eos}
\end{figure}

\begin{figure*}[htb]
    \centering
    \includegraphics[width=\textwidth]{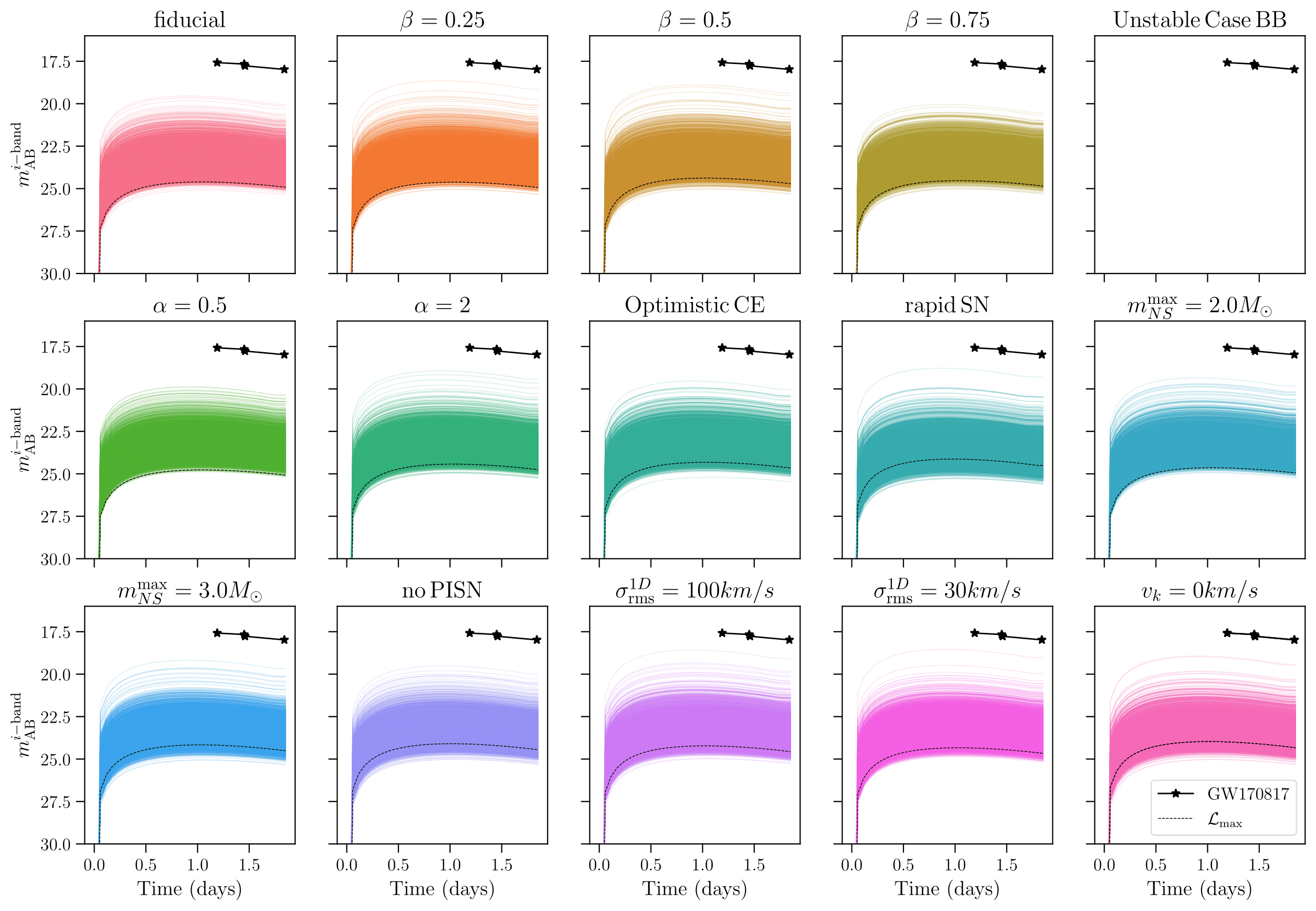}
    \caption{\acp{KN} light curves in the $i-$band corresponding to the different population models for the \texttt{DD2} EoS. The dotted line shows the light curve associated with the maximum likelihood parameter estimates for each model.}
    \label{fig:kilonovae-dd2}
\end{figure*}

\begin{figure}[htb]
    \centering
    \includegraphics[width=0.45\textwidth]{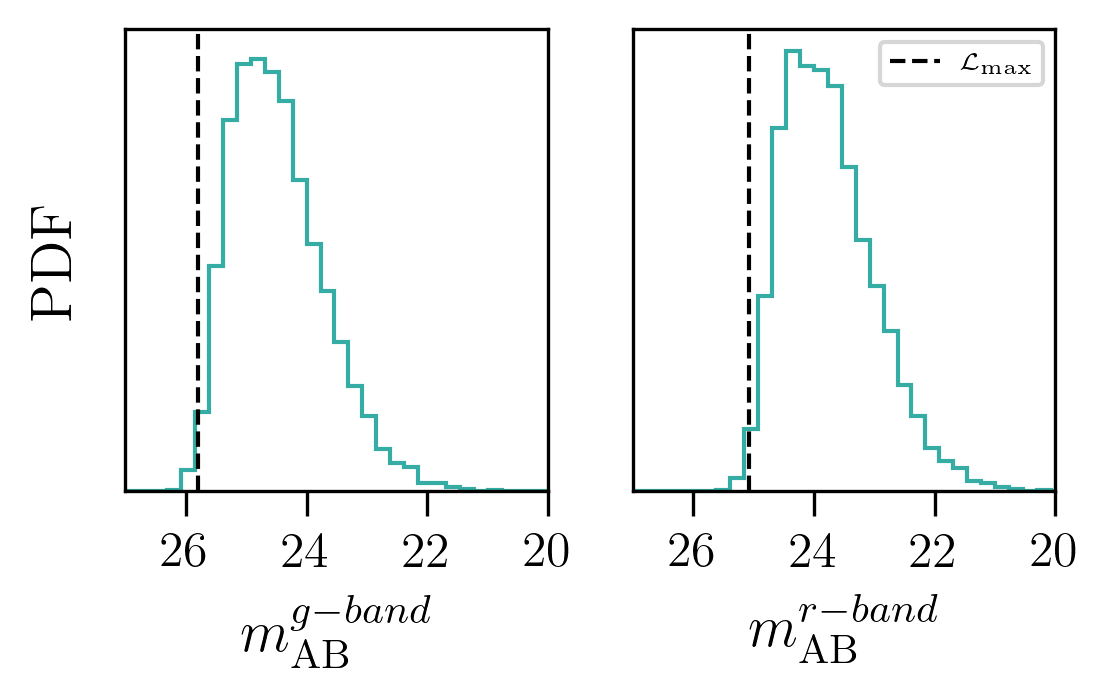}
    \caption{The distribution of the peak luminosity in the $g$ and $r$ bands for the \texttt{DD2} EoS corresponding to samples from the ``fiducial" population model. The black dotted line shows the peak luminosity corresponding to the maximum likelihood binary parameters.}
    \label{fig:peak-dd2}
\end{figure}

As noted in Sec.~\ref{sec:lightcurves}, the KNe light curves were generated using numerical recipes in \citet{Gupta:2023evt}. We only consider samples where the mass ratio is less than 4. We assume the electron fraction for the dynamical ejecta to be 0.1 and for the unbound disk ejecta to be 0.3 \citep{Ekanger:2023mde}. The opacity values are obtained by fitting to data from \citet{Tanaka:2019iqp}. 

The four chosen \acp{EoS} in Sec.~\ref{sec:lightcurves} span a wide range in the mass-radius diagram (see Fig.~\ref{fig:eos}). 
This allows us to gauge the effect of different nuclear matter compositions on the expected electromagnetic counterpart. Notably, the \texttt{BSK24} \ac{EoS}~\citep{PhysRevC.88.061302} -- used by~\cite{LVK:2024elc} -- is contained within this range. Fig.~\ref{fig:kilonovae-dd2} shows the results corresponding to the \texttt{DD2} \acp{EoS}, which show similar features to those from \texttt{H4}.

To compare our results with \citet{Zhu:2024cvt}, we, in Fig.~\ref{fig:peak-dd2}, show the distribution of the peak luminosity in the $g$ and $r$ bands for the ``fiducial" model and the \texttt{DD2} \ac{EoS}. While the two works get similar estimates for peak luminosities, our model results in dimmer KNe compared to the one used by \citet{Zhu:2024cvt}. Furthermore, the maximum likelihood binary parameters correspond to a significantly fainter KNe, affecting the KNe's detectability from such systems.

\bibliography{sample631}{}
\bibliographystyle{aasjournal}

\end{document}